\documentclass[epj,final]{svjour}
\usepackage{graphics,epsfig,rotating}
\hugehead

\newcommand{\be}{\begin{equation}}
\newcommand{\ee}{\end{equation}}
\newcommand{\bea}{\begin{eqnarray}}
\newcommand{\eea}{\end{eqnarray}}
\begin{document}

\title{Two-proton small-angle correlations in central heavy-ion collisions: a beam-energy 
and system-size dependent study}
\subtitle{\rm FOPI Collaboration}
\author{\rm
R.~Kotte\inst{5}
\and J.~P.~Alard\inst{3} 
\and A.~Andronic\inst{1,4} 
\and V. Barret\inst{3}
\and Z.~Basrak\inst{12} 
\and N.~Bastid\inst{3}  
\and M.L.~Benabderrahmane\inst{6}
\and R.~\v{C}aplar\inst{12} 
\and E.~Cordier\inst{6} 
\and P.~Crochet\inst{3} 
\and P.~Dupieux\inst{3} 
\and M.~D\v{z}elalija\inst{12} 
\and Z.~Fodor\inst{2} 
\and I.~Ga\v{s}pari\'{c}\inst{12}
\and A.~Gobbi\inst{4} 
\and Y.~Grishkin\inst{7}
\and O.N.~Hartmann\inst{4}
\and N.~Herrmann\inst{6} 
\and K.D.~Hildenbrand\inst{4} 
\and B.~Hong\inst{9} 
\and J.~Kecskemeti\inst{2} 
\and Y.J.~Kim\inst{9} 
\and M.~Kirejczyk\inst{6,11}    
\and P.~Koczon\inst{4} 
\and M.~Korolija\inst{12}  
\and T.~Kress\inst{4} 
\and A.~Lebedev\inst{7}  
\and Y.~Leifels\inst{6} 
\and X.~Lopez\inst{3}
\and M.~Merschmeyer\inst{6} 
\and J.~M\"osner\inst{5}
\and W.~Neubert\inst{5}  
\and D.~Pelte\inst{6} 
\and M.~Petrovici\inst{1} 
\and F.~Rami\inst{10} 
\and W.~Reisdorf\inst{4} 
\and B.~de Schauenburg\inst{10}  
\and A.~Sch\"uttauf\inst{4} 
\and Z.~Seres\inst{2} 
\and B.~Sikora\inst{11} 
\and K.S.~Sim\inst{9} 
\and V.~Simion\inst{1} 
\and K.~Siwek-Wilczy\'{n}ska\inst{11} 
\and V.~Smolyankin\inst{7} 
\and G.~Stoicea\inst{1} 
\and Z.~Tyminski\inst{4,11}
\and P.~Wagner\inst{10} 
\and K.~Wi\'{s}niewski\inst{11} 
\and D.~Wohlfarth\inst{5} 
\and Z.G.~Xiao\inst{4}
\and Y.~Yushmanov\inst{8} 
\and A.~Zhilin\inst{7} 
}
\institute{\renewcommand{\thefootnote}{{\rm\alph{footnote}}}
\setcounter{footnote}{0}
Institute for Nuclear Physics and Engineering, Bucharest, Romania 
\and
Central Research Institute for Physics, Budapest, Hungary 
\and
Laboratoire de Physique Corpusculaire, IN2P3/CNRS and Universit\'e Blaise Pascal, 
Clermont-Ferrand, France
\and
Gesellschaft f\"ur Schwerionenforschung, Darmstadt, Germany 
\and
IKH, Forschungszentrum Rossendorf, PF 510119, D-01314 Dresden, Germany,  
\email{kotte@fz-rossendorf.de}
\and
Physikalisches Institut der Universit\"at Heidelberg, Heidelberg, Germany 
\and
Institute for Theoretical and Experimental Physics, Moscow, Russia
\and
Kurchatov Institute for Atomic Energy, Moscow, Russia 
\and
Korea University, Seoul, Korea
\and
Institut de Recherches Subatomiques, IN2P3-CNRS/ULP, 
Strasbourg, France 
\and
Institute of Experimental Physics, Warsaw University, Warsaw, Poland 
\and
Rudjer Bo\u skovi\'c Institute Zagreb, Zagreb, Croatia 
}
\date{Received: \today}
\titlerunning{Two-proton small-angle correlations in central heavy-ion collisions} 
\authorrunning{R.~Kotte {\it et al.}}
\abstract{
Small-angle correlations of pairs of protons emitted in central collisions of Ca + Ca, 
Ru + Ru and Au + Au at beam energies from 400 to 1500 MeV per nucleon are 
investigated with the FOPI detector system at SIS/GSI Darmstadt.   
Dependences on system size and beam energy are presented which extend the 
experimental data basis of pp correlations in the SIS energy range substantially. 
The size of the proton-emitting source is estimated by comparing the experimental 
data with the output of a final-state interaction model which utilizes either   
static Gaussian sources or the one-body phase-space distribution of protons provided by 
the BUU transport approach. The trends in the experimental data, {\it i.e.} system-size 
and beam energy dependences, are well reproduced by this hybrid model. 
However, the pp correlation function is found rather insensitive to the 
stiffness of the equation of state entering the transport model calculations.}


\maketitle

\section{Introduction} 
\label{intro}
Two-proton correlation functions at small relative momenta 
can probe the space-time extent of the reaction zone created in energetic 
heavy-ion reactions. This is due to the fact that the magnitude of nuclear 
and Coulomb final-state interactions (FSI) as well as anti-symmetrization effects  
depend on the spatial separation of the two protons during the emission process 
\cite{Koonin77,Pratt,Pochod,Dupieux,Gong,Bauer1,Bauer2,Bauer3,Kunde93,Lisa1,Lisa2,Lisa3,Handzy1,Handzy2,Handzy3,Martin96,Fritz99,Schwarz01,Kotte97,Kotte98,Kotte99}. 
Usually, the correlation functions are evaluated
as a function of the magnitude q of the relative momentum vector 
${\bf q}=({\bf p}_1 - {\bf p}_2)/2$. The interplay of the 
attractive S-wave nuclear interaction and the Coulomb repulsion and anti-symmetrization 
produce a minimum at q=0 and a maximum in the correlation function at 
$q \sim 20$~MeV/c \cite{Koonin77}.
Most analyses give only upper limits for the spatial extent of the source 
due to the ambiguity of radius and lifetime of the source, {\it i.e.} 
model calculations simulating large sources with short lifetimes will give 
very similar correlation functions as model calculations simulating small 
sources with long lifetimes \cite{Koonin77,Gong}. 

At beam energies below about 100 $A\cdot$MeV, heavy-ion experiments performed 
mainly at the National Superconducting Cyclotron Laboratory (NSCL) at Michigan State University 
(MSU) \cite{Pochod,Gong,Lisa1,Lisa2,Lisa3,Handzy1,Handzy2,Handzy3,Martin96} 
allowed for systematic investigations of pp small-angle correlations. Furthermore,  
new methods allowing to deduce the emission source function from two-particle correlations 
have been proposed. Thus, recently, the technique of source imaging 
\cite{Brown97,Brown98,Panitkin00,Brown00}, {\it i.e.} the numerical inversion  
of the correlation function, has been applied successfully to pp correlations 
studied in heavy-ion experiments 
not only in the MSU-NSCL energy range \cite{Brown97,Verde02,Verde03}  
but also at significantly higher beam energies provided by the Brookhaven-AGS \cite{Chung03} 
or even by the CERN-SPS \cite{BrownWang99}. 

However, in the energy domain of the heavy-ion synchrotron SIS at GSI Darmstadt, 
ranging from about 100 to 2000 $A\cdot$MeV, 
only few data on pp correlations are available: 
proton-proton correlations at small relative momenta were measured with the  
``Plastic Ball'' at BEVALAC for two systems, Ca on Ca and Nb on Nb, 
at 400 $A\cdot$MeV as a function of proton multiplicity \cite{Gustafson84}. 
Freeze-out densities of about 25\% of normal nuclear-matter density were extracted. 
Two-proton small-angle correlations were studied 
at SATURNE in Ar + Au reactions at 200 $A\cdot$MeV \cite{Kunde93} and   
in Ne- and Ar-nucleus collisions between 200 and 1000 $A\cdot$MeV \cite{Dupieux}. 
Breakup densities and time scales in spectator fragmentation following the reaction 
Au + Au at 1 $A\cdot$GeV were investigated with the ALADiN spectrometer at SIS/GSI 
\cite{Fritz99,Schwarz01}. Finally, the space-time extent of the proton emitting source 
was studied  with the $4\pi$ detector FOPI measuring 
central collisions of Ni + Ni at 1930 $A\cdot$MeV and of Ru + Ru at 
400 $A\cdot$MeV  \cite{Kotte97,Kotte99}. Consequently, the 
extension of the pp-correlation data base is highly desirable. 

Since the theoretical pp correlation function only requires the knowledge of the 
one-body phase-space distribution of protons, it is possible to generate small-angle 
correlations with any microscopic theory that calculates the time evolution 
of the one-body distribution function \cite{Bauer2}. In addition to the usage of 
Gaussian source distributions we will also follow this procedure 
since it allows a more adequate comparison of experimental and theoretical correlation 
functions. 
Microscopic transport models, like Boltzmann-Uehling-Uhlenbeck (BUU) \cite{Bauer1,Bauer2,Bauer3} 
or Quantum Molecular Dynamics (QMD) \cite{Bass95} in their various implementations, 
do not only allow to trace the expectation values of the 
positions of the various particles in coordinate space 
but also the corresponding time information. Moreover, they carry the  
correlation of coordinate and momentum space coordinates of the individual particles. 
This correlation is generated 
intrinsically during the expansion following the compression phase 
of the heavy-ion collision. On a macroscopic scale, the expansion constitutes a 
collective motion of the various particle species, called radial flow. Thus, it is 
natural to expect that a hybrid model consisting of BUU approach plus FSI model    
should be able to reproduce the experimental data and give the relevant insight into the 
physics of the heavy-ion collision until particle freeze out. 
Furthermore, there is reasonable hope that a systematic comparison of as much as possible 
experimental information, 
{\it e.g.} on sideward \cite{Andronic03} and elliptic flow \cite{Stoicea04}, 
on nuclear stopping \cite{Reisdorf041}
and cluster production \cite{Reisdorf042}, on the shape (in momentum space) 
of the participant source \cite{Bastid04} or on charged pion production \cite{Hong02}, 
with transport model predictions will allow to constrain the essential 
input parameters of the model, {\it e.g.} the equation of state (EoS) and/or the  
nucleon-nucleon cross section within the nuclear medium. 
Earlier it was reported that the pp correlation function generated for $^{14}$N+$^{27}$Al 
collisions at 75~$A\cdot$MeV is only weakly dependent on the EoS \cite{Gong,Bauer3}. 
It is not clear whether the sensitivity to the microscopic details of the transport approach 
increases for higher beam energies and/or heavier projectile-target combinations. 
Therefore, in the following we will study in a systematic way 
how the BUU+FSI hybrid model reproduces the data and what kind of information can be gained. 

The paper is structured as follows: In sect.\,\ref{experiment} the experimental 
basis is described shortly, the involved event classes are explained and 
the correlation function is defined. 
In sect.\,\ref{results} the experimental pp correlation functions 
with their dependences on system size and beam energy 
are presented and compared to model predictions. 
Finally, the results are summarized in sect.\,\ref{summary}.

\section{The experiment}  \label{experiment}
The experiments have been performed at the heavy-ion synchrotron SIS at GSI
Darmstadt. Symmetric collisions are carried out by irradiating 
targets of 1~\% interaction thickness of 
$^{40}$Ca and $^{197}$Au with the corresponding ions of beam  
energies between 400 and 1500 $A\cdot$MeV. At a projectile energy of 400 $A\cdot$MeV 
the system of intermediate mass, $^{96}$Ru + $^{96}$Ru, is included which was 
explored extensively with respect to the 
space-time difference of light charged particles emission in central collisions \cite{Kotte99}.  

The present analysis uses a subsample of the data, 
taken with the outer Plastic Wall/Helitron 
combination of the FOPI detector system \cite{Gobbi93,Ritman95}. 
The Plastic Wall delivers - via energy loss vs. time-of-flight (TOF) 
measurement - the nuclear charge $Z$ and the velocity $\beta$ 
of the produced particles and the corresponding hit positions. 
The Helitron gives the curvature (which is a measure 
of momentum over charge $(p/Z)$)   
of the particle track in the field of a large super-conducting solenoid.
Since the momentum resolution of the Helitron is rather moderate, 
this detector component serves for particle identification only. The  
mass $m$ is determined 
via  $m c=(p/Z)_{Hel}/(\beta \gamma/Z)_{Pla}$, where 
$\gamma=(1-\beta^2)^{-1/2}$. 
The Plastic Wall and the Helitron have full overlap only for 
polar angles between 8.5 degrees and 26.5 degrees. 
The corresponding flight paths amount to 450~cm and 380~cm, 
respectively. 
Monte-Carlo simulations have been performed
in order to study the influence of the finite detector 
granularity and of the TOF and position resolutions  
on the velocity and finally on the proton momentum. 
The resolution of both quantities is governed by the TOF resolution, which 
is $\sigma_{TOF} = 80$~(120)~ps for short (long) scintillator 
strips located at small (large) polar angles \cite{Gobbi93}. 
The velocity can be determined with a precision 
of $\delta \beta/\beta <1$~\%. 

\subsection{Event classification} \label{centrality}
Typically, a few hundred thousand to a million central events are collected 
for each individual collision system 
by demanding large charged-particle multiplicities to be measured 
in the outer Plastic Wall. The corresponding 
integrated cross sections for the collision systems Ca + Ca, Ru + Ru, and Au + Au
comprise about 15~\%, 10~\% and 10~\% of the total cross section, respectively. 
For the present central-event class one would expect - within a geometrical picture - 
average impact parameters of about 2~fm, 2~fm, and 3~fm. Taking into account typical 
dispersions of the impact parameter distribution of about 1~fm as found from 
GEANT simulations \cite{GEANT94}, the number of participants in the central source 
is estimated to $A_{part}=56\pm12~(147\pm21,~302\pm34)$  
for the system Ca + Ca (Ru + Ru, Au + Au). 
More peripheral collisions have been measured, 
too. But, due to both the strong downscaling (typically a factor of 16) 
of the corresponding online trigger and the lower proton multiplicities, the 
pp pair statistics was not sufficient for correlation analyses. 

In previous investigations of central Au + Au collisions between 100 and 
400 $A\cdot$MeV beam energy it was found that the  
correlation function of pairs of intermediate mass fragments  
is strongly affected by the collective directed sideward flow of nuclear
matter \cite{Kaempfer93,Kotte95}. This directed 
sideflow causes an enhancement of correlations at small relative
momenta. The enhancement results from mixing of differently
azimuthally oriented events; it vanishes if the events are rotated into a 
unique reaction plane, which is determined by the standard   
transverse momentum analysis \cite{Odyniec}. 
Consequently, the technique of event rotation is applied also 
to the present data in order to prevent that such artificial correlations  
are introduced into the reference momentum distribution of the 
correlation function (cf. sect.\,\ref{corr_fun}). 

\subsection{Correlation function}\label{corr_fun}
Let $Y_{12}({\bf p}_1, {\bf p}_2)$ be the coincidence yield of pairs 
of particles having momenta ${\bf p}_1$ and ${\bf p}_2$. 
Then the two-particle correlation function is defined as 
\begin{equation}
1 + \mbox{R}({\bf p}_1, {\bf p}_2) = {\cal N} \,
\frac{\sum _{events,pairs} Y_{12}({\bf p}_1, {\bf p}_2)}
{\sum_{events,pairs} Y_{12,mix}({\bf p}_1, {\bf p}_2)}.
\label{def_exp_corr_fct}
\end{equation}
The sum runs over all events fulfilling the above mentioned global 
selection criterion and over all pairs satisfying certain conditions given 
below. 
Event mixing, denoted by the subscript ''mix'',
means to take particle 1 and particle 2 from different events.
The normalization factor ${\cal N}$ is fixed by the requirement to have the
same number of true and mixed pairs. 
The statistical errors of all the correlation functions presented below 
are governed by those of the coincidence yield, since the mixed yield is
generated with two orders of magnitude higher statistics. 
The correlation function (\ref{def_exp_corr_fct}) is then projected onto the relative momentum 
\begin{equation}
q = \vert {\bf q} \vert= \frac{1}{2} \, \vert {\bf p}^{cm}_1 - {\bf p}^{cm}_2 \vert.
\label{defq}
\end{equation}
Here, ${\bf p}^{cm}_{i}$ are the particle momenta calculated 
in the c.m. system of the colliding nuclei.
From the velocity resolution as estimated in 
sect.\,\ref{experiment} the corresponding $q$ resolution is 
deduced to be $\delta q=(4\pm 1)$~MeV/c. A similar value can be derived directly from 
the experimental dispersion of the resonance due to the narrow 2.186~MeV state of $^6$Li 
($J^{\pi}=3^+$, $\Gamma=24$~keV) in the $\alpha$-d  
correlation function \cite{Kotte99}. All theoretical correlation functions presented 
in sect.\,\ref{results} are folded with this $q$ resolution. 

As in previous proton-proton correlation analyses \cite{Kotte97,Kotte98,Kotte99} an 
enhanced coincidence yield 
at very small relative angles is observed, which is  
due to double counting caused 
mainly by scattering in the scintillator strips. This disturbing yield  
is reduced strongly 
by the requirement to match the particle hits on the Plastic Wall 
with the corresponding tracks in the forward drift chamber Helitron. 
The remaining left-over of doubly counted scattered particles
is eliminated by excluding, around a given hit,
positions within a rectangular segment of azimuthal and polar angle 
differences  
$\vert \phi_1 -\phi_2 \vert <4^o$ and $\vert \theta_1 -\theta_2 \vert <2^o$.
The same procedure is applied to the uncorrelated background, hence keeping the 
influence of the exclusion onto the correlation function as small as possible. 
GEANT simulations \cite{GEANT94} have shown that at very small 
relative momenta, $q <(12-15)$~MeV/c, still a small 
bias of the correlation function can not be excluded \cite{Kotte97,Kotte99}. 
Thus, the corresponding regions in the correlation functions are not
taken into consideration when comparing the experimental data with model 
predictions. 

\section{Results and comparison with model predictions} \label{results}

For description of the FSI model we refer to ref.\,\cite{Kotte99} and 
references cited therein. As source distributions 
we use either Gaussian density profiles in coordinate 
and momentum space or the output of the BUU transport model. 
The theoretical correlation function is given as 
\begin{eqnarray}
1+R({\bf P},{\bf q}) &=& \int d^3r\,S({\bf r},{\bf P})\,\vert \Psi_{\bf q}({\bf r})\vert^2.
\end{eqnarray}
Here, the wave function $\Psi$ describes the relative motion of the two emitted particles and 
$S$ is the source function. Gaussian sources are defined in \cite{Kotte99}. In case of  
calculating the single-particle phase-space distribution within the BUU model  
the source function is given as 
\begin{eqnarray}
\nonumber
S({\bf r},{\bf P}) &=&
{\cal N}_p \sum_{i,j}\,\delta(\frac{{\bf P}}{2}-{\bf p}_i)\,\delta(\frac{{\bf P}}{2}-{\bf p}_j) \times \\
&&\delta({\bf r}-({\bf r}_i-{\bf r}_j)+\frac{{\bf P}}{2 m}(t_i-t_j)). 
\end{eqnarray}
${\bf r}_i$, $t_i$, and ${\bf p}_i$ are the space, time and momentum coordinates 
of particle $i$ at freeze out.  
The parameter ${\bf P}$ represents the sum momentum of the observed particle pair. 
The normalization ${\cal N}_p$ is chosen such that 
\begin{equation}
\int d^3r\,S({\bf r},{\bf P})=1.
\end{equation} 
To allow the summation over a sufficient number of particles, the  
$\delta$ function treating the momentum is replaced by a Gaussian with dispersion $\Delta_p$, 
\begin{equation} 
\delta({\bf p}) \rightarrow  \bigl(\frac{1}{2 \pi \Delta_p^2}\bigr)^{3/2} \exp\bigl(-\frac{{\bf p}^2}{2 \Delta_p^2}\bigr).
\end{equation} 
Note that all coordinates are calculated in the c.m. system of the colliding nuclei.
The relevant BUU input parameters are
i) the number of test particles per nucleon ({\it i.e.} parallel runs or pseudo events, 
here chosen to 200),
ii) nuclear charge and mass numbers of target and projectile, $A_t$, $Z_t$, $A_p$, $Z_p$,
iii) the projectile energy $E_{proj}$, 
iv) the impact parameter $b$, and 
v) the stiffness of the Equation of State (EoS), either hard ($\kappa=380$~MeV), medium 
($\kappa=290$~MeV) or soft ($\kappa=215$~MeV) \cite{Daffin95}. 
A proton is considered ``emitted'' when the surrounding density falls below a value of
$\rho_{cutoff}=0.02$/fm$^3$ and when subsequent interaction with the mean field does not 
cause recapture into regions of higher density until the 
calculation is terminated at 150~fm/c. 
The cutoff value of about one eighth of normal 
nuclear matter density implies ceasing NN interactions in a sufficiently diluted nuclear matter. 
This choice was approved in systematic investigations of proton-proton 
correlations in $^{14}$N+$^{27}$Al and $^{14}$N+$^{197}$Au reactions at 
75~$A\cdot$MeV \cite{Gong}, 
in $^{36}$Ar+$^{45}$Sc collisions at 80~$A\cdot$MeV \cite{Lisa1,Handzy1}   
and in $^{40}$Ar+$^{197}$Au reactions at 200~$A\cdot$MeV \cite{Kunde93}.    

Finally, for each run and each proton, the septuplet of space-time-momentum coordinates 
$N_i=(x,y,z,t,p_x,p_y,p_z)_i$ is taken from the BUU approach at the 
corresponding freeze-out time and then processed within the FSI model. 
The relevant parameters of the FSI model are the 
so-called ``observation quantities'', {\it i.e.} the sum momentum $P$, the momentum dispersion  
$\Delta_p$ and the polar angle $\theta^{cm}$. These quantities are taken 
directly from the experimental distributions in the c.m. system, {\it e.g.} 
$P=2\,\langle p^{cm} \rangle$.  Typical values are  $\theta^{cm}=45^o$,  
$\Delta_p=150$~MeV/c, and $\langle p^{cm} \rangle=300~(400)$~MeV/c for 400 (1500) $A\cdot$MeV
beam energy. 

\begin{figure}[h]
\centering
\mbox{
\epsfxsize=1.\linewidth
\epsffile{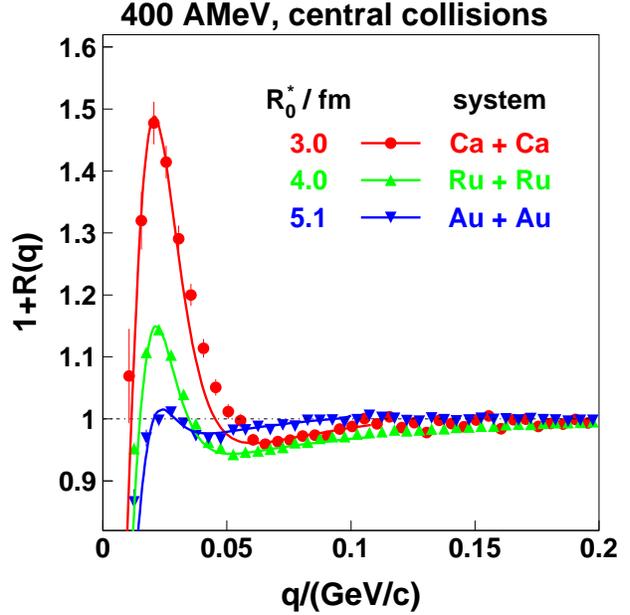} 
     }
\caption{
Correlation functions of proton pairs from central collisions of Ca + Ca, 
Ru + Ru, and Au + Au at 400 $A\cdot$MeV. 
Experimental data (symbols) are compared to predictions of the 
FSI model with Gaussian sources and zero lifetime (lines).  
The corresponding apparent source radii are indicated. 
\label{pcor_vs_system_400_hwb}
        }
\end{figure}

\begin{figure}[h]
\centering
\mbox{
\epsfxsize=1.\linewidth
\epsffile{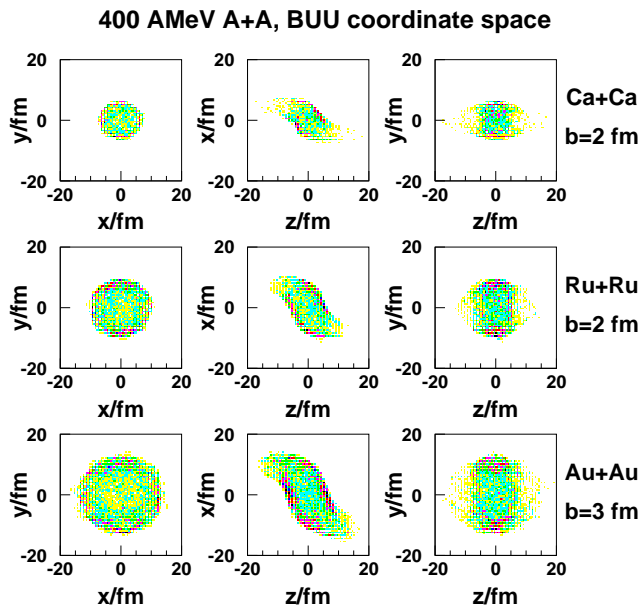} 
     }
\caption{
Coordinate-space distributions of freeze-out points   
of protons from central collisions of Ca + Ca (top), 
Ru + Ru (center), and Au + Au (bottom) at 400 $A\cdot$MeV simulated   
by the BUU transport approach. The corresponding impact parameters are indicated at 
the right margin. The left, middle and right columns give the 
distributions in the $(y,x)$, $(x,z)$ and $(y,z)$ plane, respectively. 
\label{coord_distr_buu_ca_ru_au}
        }
\end{figure}

\begin{figure}[h]
\centering
\mbox{
\epsfxsize=1.\linewidth
\epsffile{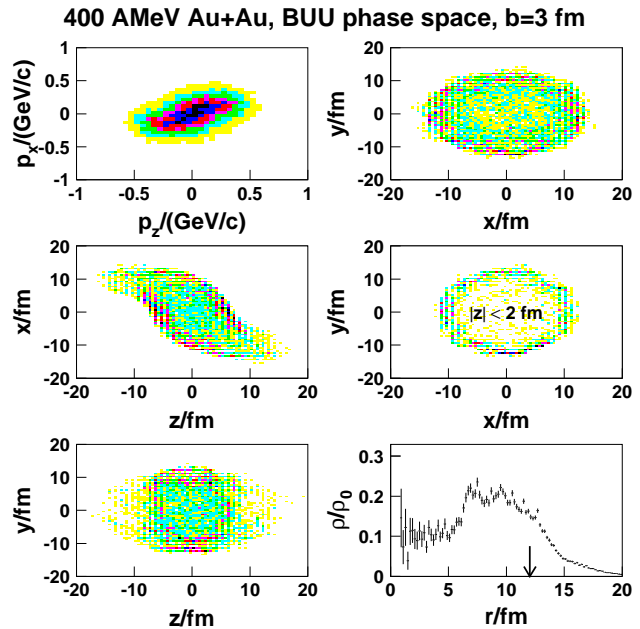} 
     }
\caption{
Phase-space distribution of freeze-out points of 
protons from central collisions of Au + Au at 400 $A\cdot$MeV 
simulated by the BUU transport approach for an impact parameter of $b=3$~fm.  
Left column (from top to bottom): The distributions in the $(p_x,p_z)$,  
$(x,z)$, and $(y,z)$ plane. 
Right column: The $(y,x)$ distribution in total (top) and for a central $z$ slice, 
$\vert z \vert < 2$~fm (center). The bottom panel 
gives the corresponding radial density profile (in units of the saturation density). 
The arrow indicates the r.m.s. radius of the source.  
\label{phase_space_distr_buu_auau400}
        }
\end{figure}
\begin{figure}[h]
\centering
\mbox{
\epsfxsize=1.\linewidth 
\epsffile{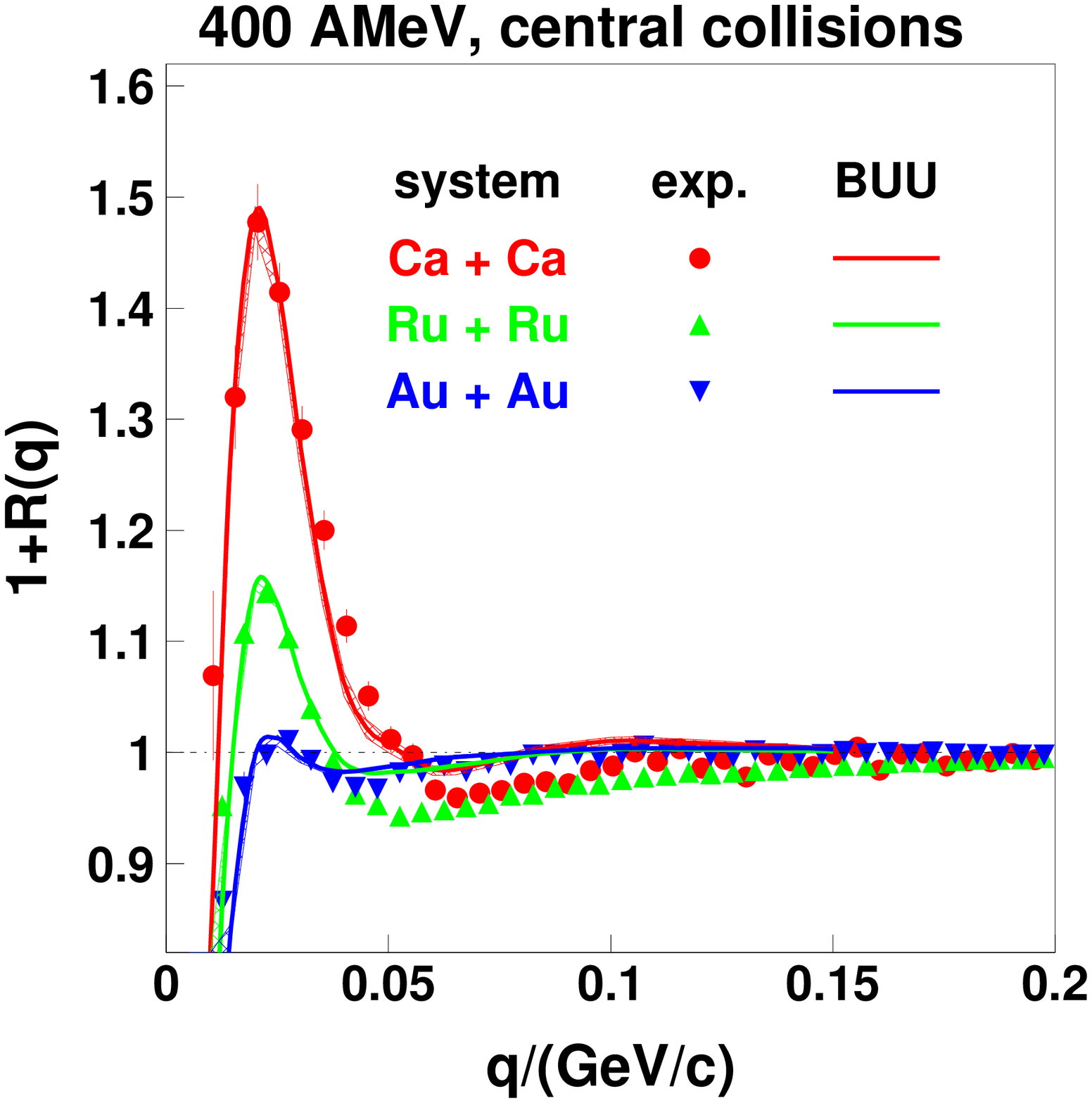} 
     }
\caption{
The same as fig.\,\ref{pcor_vs_system_400_hwb}, but here the experimental 
correlation functions (symbols) are compared to predictions of the 
BUU+FSI model (lines). The narrow 
hatched bands around the lines indicate the systematic variation of 
the calculations for the three different parameterizations of the EoS (see text).
\label{pcor_vs_system_400_buu}
        }
\end{figure}
 
\subsection{System-size dependence}\label{sys_size_dep}
Figure\,\ref{pcor_vs_system_400_hwb} shows, for 400 $A\cdot$MeV beam energy, the system-size 
dependence of the pp correlation function. Clearly, the pp-correlation peak 
decreases with increasing system size and almost vanishes for the largest collision 
system Au + Au. The peaks ({\it i.e.} data within a relative-momentum range of $15<q/$MeV/c$<50$) 
of the experimental correlation functions (symbols) are best described by the theoretical 
ones (lines) for apparent Gaussian radii of 
$R_0^*=(3.0\pm 0.1)$~fm for Ca + Ca, $(4.0\pm0.15)$~fm for Ru + Ru and $(5.1\pm 0.2)$~fm for 
Au + Au collisions. Note that the term ``apparent'' means that neither the contribution  
of the finite emission duration 
(inappropriately called ``lifetime'', leading to an increase of the effective source radius) 
nor the effect of collective radial expansion of the participant zone 
(leading - via strong correlations of coordinate and momentum space - to a reduction of the 
effective source size) are taken into account \cite{Kotte97}. In the following, 
the asterisk indicates apparent quantities, {\it e.g.} the sharp-sphere and r.m.s. radii 
$R^*_{ss}=\sqrt{5}\,R^*_0$ and $R^*_{rms}=\sqrt{3}\,R^*_0$, respectively. 
For Gaussian source distributions the effective radius can be written as \cite{Kotte99} 
\bea
R_0^* = \sqrt{\frac{R_0^2}{1+\epsilon}+(v\tau)^2},
\label{rstar}
\eea
where $\tau$ is the duration of emission, $v=P/2m$ is the pair velocity, 
and $\epsilon=\epsilon_{flow}/\epsilon_{therm}$ represents the ratio of radial flow energy 
$\epsilon_{flow}$ and the energy of the random thermal motion $\epsilon_{therm}=\frac{3}{2}T$. 

For both the small and the medium-size systems, our results can be compared directly with 
existing data: Gaussian radii $r_g$ were derived from pp 
correlations measured with the ``Plastic Ball'' at BEVALAC 
for the reactions Ca + Ca and Nb + Nb, both at 400 MeV per nucleon,  
as function of proton multiplicity \cite{Gustafson84}. Taking into account the different 
radius definitions, i.e. $r_g=\sqrt{2} R^*_0$, 
the apparent source radii for central collisions agree perfectly. 
Consequently, the freeze-out density of about 25\% of normal nuclear matter density  
deduced in ref.\,\cite{Gustafson84} is in accordance with our results presented 
in sect.\,\ref{beam_dep} below.  
Instead of disentangling the space-time-flow ambiguity of 
eq.\,(\ref{rstar}) \cite{Kotte97,Kotte99}, 
in the following we try to reproduce the experimental correlation 
functions by BUU+FSI model calculations.

Before we confront the data with theoretical correlation functions 
we briefly focus onto the coordinate-space distributions 
of freeze-out points from BUU calculations displayed in 
fig.\,\ref{coord_distr_buu_ca_ru_au} for the three systems 
at 400 $A\cdot$MeV beam energy. 
Note that the coordinates are not taken at fixed time but when the 
local density drops below the cutoff value. 
As expected, the proton source increases with increasing system size. 
Additionally to the coordinate-space distribution shown in 
fig.\,\ref{coord_distr_buu_ca_ru_au}, as an example for the Au + Au system  
fig.\,\ref{phase_space_distr_buu_auau400} displays the momentum 
distribution projected onto the reaction plane $(p_x,p_z)$, the $(y,x)$ distribution  
for a central $z$ slice and the radial density profile. 
Here, the effective density is given in units of normal nuclear matter density,  
$\rho_0=0.168/$fm$^3$. The corresponding r.m.s. radius 
$R_{rms}=\sqrt{\langle r^2\rangle}$ of the proton emitting source is indicated by an arrow.
No essential changes 
are observed from these phase-space plots when increasing the beam energy,  
except a trivial extension in momentum space. Since even the r.m.s. radius
of the breakup configuration (cf. arrow in lower right panel) 
is almost the same for 400 and 1500 $A\cdot$MeV beam energy, 
one would expect a very similar height of the pp correlation peaks. However, also  
the correlation of coordinate and momentum space, constituting a fast collective motion known as 
radial expansion, is expected to increase with 
beam energy. It should lead to a reduction of the apparent source radius and hence to 
an increase of the pp peak height (cf. eq.\,(\ref{rstar})). This presumption will be verified 
in sect.\,\ref{beam_dep}. 

For 400 $A\cdot$MeV beam energy, fig.\,\ref{pcor_vs_system_400_buu} compares the 
experimental system-size dependence of the pp correlation function with the 
corresponding output of the BUU+FSI model. 
Hardly any differences have been found between the pp correlation functions derived from BUU 
simulations using the three different parameterizations of the EoS as can be inferred from the 
narrow hatched bands in Fig.\,\ref{pcor_vs_system_400_buu}. (Similar findings 
have been reported for $^{14}$N+$^{27}$Al collisions at 75~$A\cdot$MeV \cite{Gong,Bauer3} and for 
$^{40}$Ar+$^{197}$Au reactions at 200~$A\cdot$MeV \cite{Kunde93}.)   
Therefore, in the following we will restrict ourselves to the medium stiff EoS. 

\subsection{Beam-energy dependence}\label{beam_dep}
\begin{figure}[h]
\centering
\mbox{
\epsfxsize=1.\linewidth
\epsffile{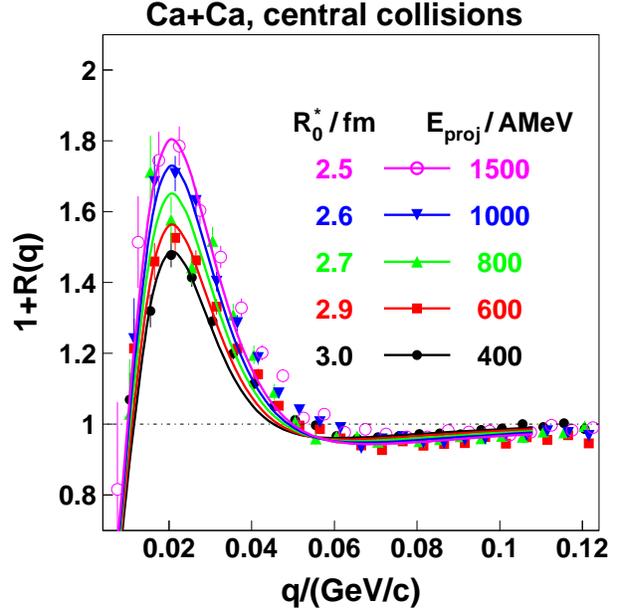} 
     }
\caption{
Correlation functions of proton pairs from central collisions of Ca + Ca 
at projectile energies from 400 to 1500 $A\cdot$MeV. 
Experimental data (symbols) are compared to predictions of the 
FSI model with Gaussian source and zero lifetime (lines).  
The corresponding apparent source radii are indicated.
\label{pcor_caca_vs_ebeam_hwb}
        }
\end{figure}
\begin{figure}[h]
\centering
\mbox{
\epsfxsize=1.\linewidth
\epsffile{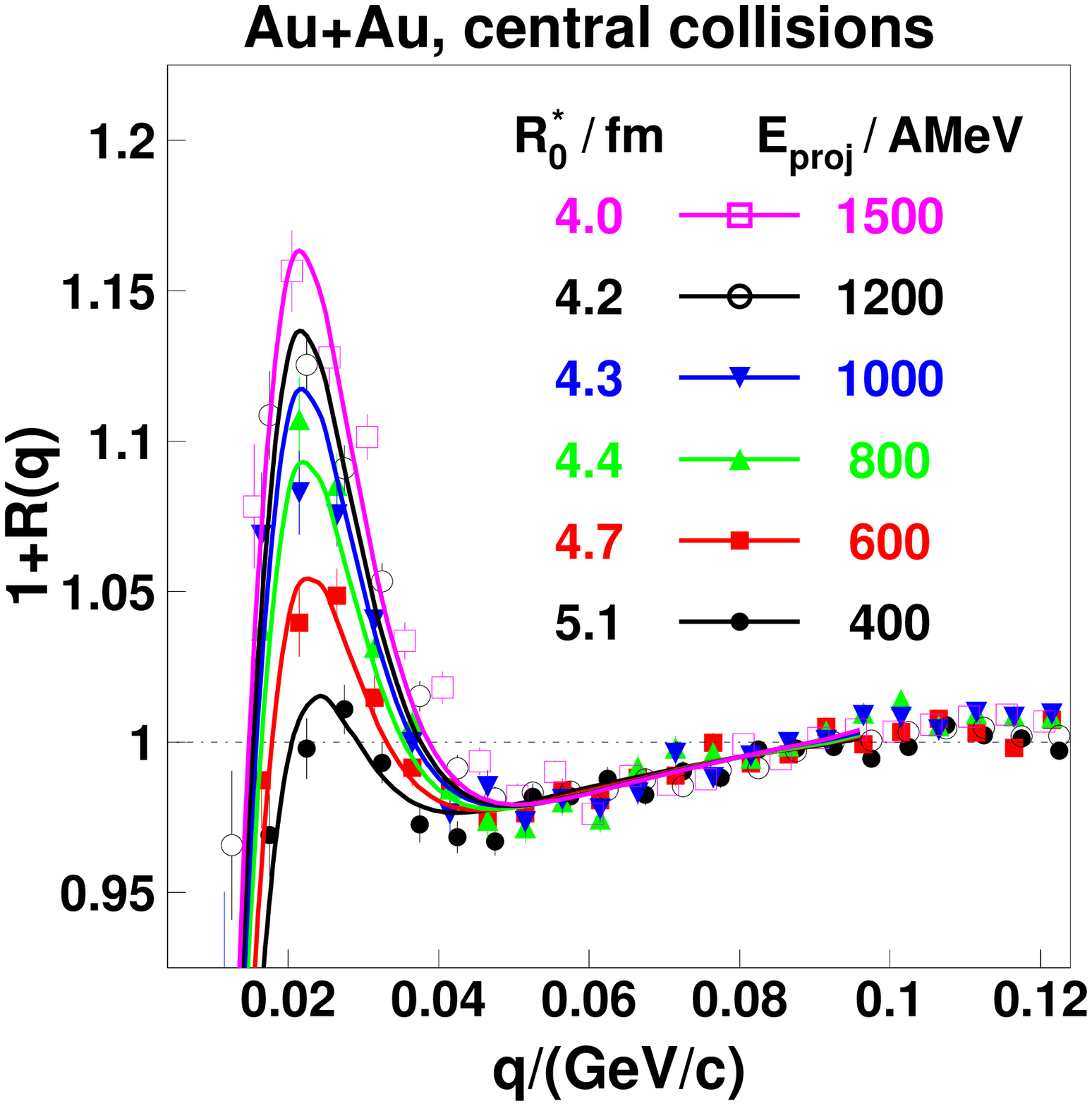} 
     }
\caption{
The same as figure\,\ref{pcor_caca_vs_ebeam_hwb}, but for Au + Au collisions. 
\label{pcor_auau_vs_ebeam_hwb}
        }
\end{figure}

Figure\,\ref{pcor_caca_vs_ebeam_hwb} shows the excitation function of experimental 
pp correlations from central collisions of Ca + Ca at beam energies from 
400 to 1500 $A\cdot$MeV (symbols). A strong increase of the correlation peak is obvious 
corresponding to a decrease of the apparent source size. Indeed, when comparing with the 
theoretical correlation functions derived from the FSI model with static Gaussian sources 
and zero lifetime (lines), the experimental 
pp peak is best reproduced with apparent radii systematically decreasing 
from 3.0~fm at 400 $A\cdot$MeV to 2.5~fm at 1500 $A\cdot$MeV. 
This shrinking of the apparent source has to be traced back to two facts.  
First of all, at higher beam energies when the heavy-ion collisions is processing faster,  
the contribution of the emission duration onto the 
effective source radius is smaller. Usually, finite emission times imply a reduced Pauli 
suppression in emission direction and hence lead to an increase of 
the effective source radius \cite{Kotte97}.) Typically, for the energy range we are 
dealing with in the present investigation, these times are in the order of 10~fm/c 
\cite{Kotte97,Kotte98,Kotte99}. 
Secondly, also the collective (radial) expansion increases with 
increasing projectile energies. It is 
well known that strong correlations of coordinate and momentum space lead to a drastic  
reduction of the apparent source radius as derived from a comparison of the data 
with results of a 
FSI model incorporating the Koonin-Pratt formalism with static Gaussian sources and finite
life (emission) times. 30~\% smaller radii are observed \cite{Kotte97,Kotte98,Kotte99} 
for a ratio of collective to random thermal energies of unity \cite{Reisdorf97}. 

Figure\,\ref{pcor_auau_vs_ebeam_hwb} shows the excitation function of experimental 
pp correlations from central collisions of Au + Au (symbols). From comparisons with the 
output of the FSI model (lines), we find an optimum reproduction of the experimental data 
in the peak region for apparent radii dropping from 5.1~fm at 400 $A\cdot$MeV to 
4.0~fm at 1500 $A\cdot$MeV. The trend found in central Au + Au collisions is  
the same as for the smaller system Ca + Ca, but with lower 
peak values. This effect is expected due to the much larger number of involved 
nucleons in central Au + Au collisions.  
Looking at the volumes $V^*=4\pi (R^*_{ss})^3/3$ derived from the 
Gaussian radii given in figs.\,\ref{pcor_vs_system_400_hwb}, 
\ref{pcor_caca_vs_ebeam_hwb} and \ref{pcor_auau_vs_ebeam_hwb}, 
our expectations are confirmed: for the same beam energy, 
the volume ratio is approximately equal to the system size ratio of 197/40. Our  
investigation now points to the breakup densities derived from the given radii. 
With the given volumes and the number of participants estimated in 
sect.\,\ref{centrality} apparent breakup densities (normalized to the density of normal 
nuclear matter) of 
$\rho^*/\rho_0=(A_{part}/V^*)/\rho_0=0.27\pm0.06~(0.29\pm0.05,~0.29\pm0.05)$ 
are deduced for Ca + Ca (Ru + Ru, Au + Au) collisions at 
400 $A\cdot$MeV. These numbers increase to values of 
$0.47\pm0.11~(0.60\pm0.10)$ for Ca + Ca (Au + Au) at 
beam energies of 1500 $A\cdot$MeV. Here, we remind once more to the fact 
that these apparent breakup densities are not corrected for both the effects of  
collective radial expansion and finite duration of emission (cf. eq.\,(\ref{rstar})). 

\begin{figure}[h]
\centering
\mbox{
\epsfxsize=1.\linewidth
\epsffile{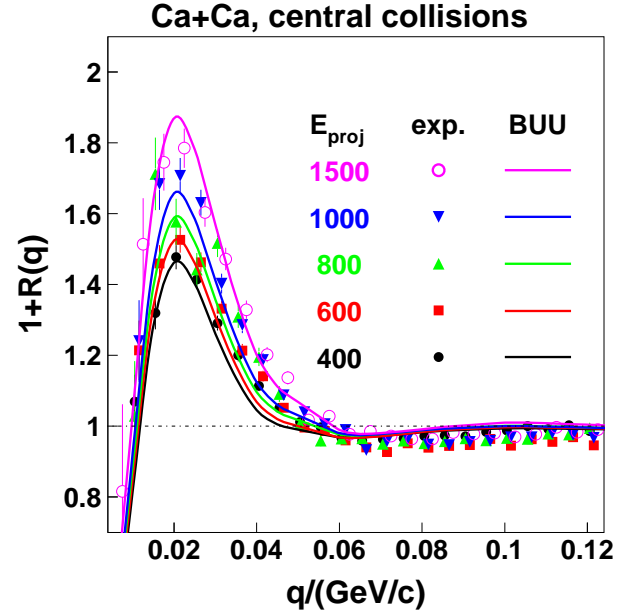} 
     }
\caption{
Correlation functions of proton pairs from central collisions of Ca + Ca 
at projectile energies from 400 to 1500 $A\cdot$MeV. 
Experimental data (symbols) are compared to predictions of the 
BUU+FSI model (lines). 
\label{pcor_caca_vs_ebeam_buu}
        }
\end{figure}

\begin{figure}[h]
\centering
\mbox{
\epsfxsize=1.\linewidth
\epsffile{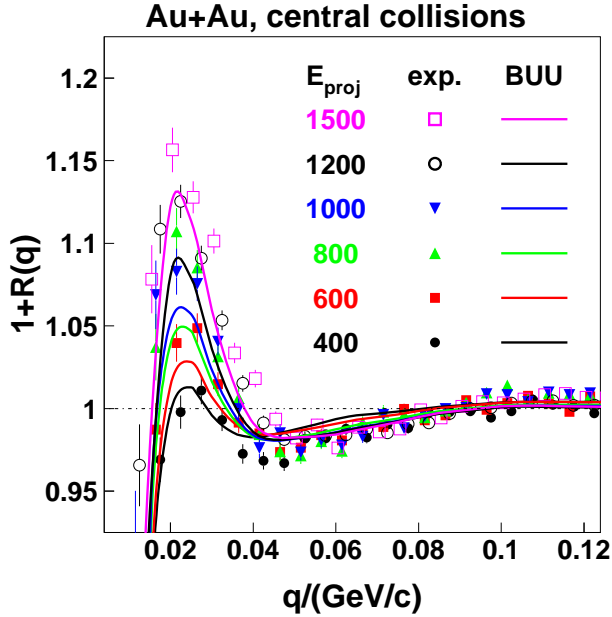} 
     }
\caption{
The same as figure\,\ref{pcor_caca_vs_ebeam_buu}, but for Au + Au collisions. 
\label{pcor_auau_vs_ebeam_buu}
        }
\end{figure}

Finally, we confront the experimental 
beam-energy dependence of two-proton correlations with the 
predictions of the BUU+FSI hybrid model.   
Figure\,\ref{pcor_caca_vs_ebeam_buu} shows the excitation function  
of pp correlations for central Ca + Ca collisions in comparison to the model results. 
With the standard parameters in the transport approach  
the excitation functions are surprisingly well reproduced. 
Also for the larger system, Au + Au, the energy dependence 
is nicely described (cf. fig.\,\ref{pcor_auau_vs_ebeam_buu}). Here, 
the experimental pp correlation peaks are systematically slightly underestimated.  

We want to point out that the agreement between our experimental 
correlation functions and those predicted by BUU is much better than in previous 
experiments performed at lower beam energies. 
Thus, pp correlation functions predicted by BUU calculations  
overestimated the measured central collision data of the reactions 
Ar + Au at  200 $A\cdot$MeV \cite{Kunde93} and of Ar + Sc at 120 and 160 
$A\cdot$MeV \cite{Handzy3}. This deficiency of the BUU model was 
attributed to its inability to treat the population of  
particle unbound resonances and their decay via delayed particle emission 
\cite{Handzy3,Verde03}. Obviously, at our higher beam energies 
this slow-emission component of the source function becomes unimportant, 
presumably since the number of heavy fragments drops 
strongly with energy \cite{Reisdorf97}. 
Thus, the fast expansion-explosion scenario seems to be 
well modelled within the transport approach. 

 
\section{Summary}\label{summary}

In the present paper, small angle correlations of pairs of protons emitted in central 
heavy-ion collisions at beam energies from 400 to 1500 $A\cdot$MeV are investigated. 
New data are presented which substantially enlarge the experimental data basis on pp 
correlations in the SIS energy range.  

The system-size and beam-energy dependences of the pp correlation peak are studied. 
The peak decreases, 
{\it i.e.} the apparent source radius increases, with increasing system size.  
This system-size  
dependence is expected from the larger volume due to a larger number of participants 
in the central source. With increasing beam energy 
the peak of the pp correlation function increases, and hence the apparent source radius 
decreases. This behaviour is interpreted as the common action of 
shorter time scales and stronger collective radial expansion. Correlations of 
coordinate and momentum space constituting the latter radial flow  
are known to be responsible for the apparent shrinking of the source size found in 
small-angle correlation studies. 

Both findings, shorter emission durations due to shorter time scales of the central  
heavy-ion collision and stronger radial expansion due to - increasing with pressure - 
space-momentum correlations, are well incorporated within 
microscopic transport models. Accordingly, using the Boltzmann-Uehling-Uhlenbeck transport 
approach, the phase-space points at freeze out are calculated and afterwards processed 
within a final-state-interaction model providing the two-proton correlation function. 
This BUU+FSI model well reproduces the system-size and beam-energy dependences of the 
experimental correlation function. However, 
the pp correlation function is found rather insensitive to the  
stiffness of the equation of state used in the transport model. 

\begin{acknowledgement}
We are grateful for many discussions to H.W.~Barz, B.~K\"ampfer and F.~Dohrmann. 
\end{acknowledgement}

\end{document}